\documentclass[useAMS,usenatbib,usegraphicx]{mn2e}

\usepackage{amssymb}
\usepackage{times}
\newcommand{\kms}{\rmn{km~s^{-1}}}
\newcommand{\msun}{\rmn{M}_{\sun}}

\title[R144 : a massive binary likely ejected from R136]
{R144 : a very massive binary likely ejected from R136 through a binary-binary encounter}

\author[S. Oh, P. Kroupa \& S. Banerjee]{Seungkyung Oh,$^1$\thanks{E-mail: skoh@astro.uni-bonn.de (SO); 
pavel@astro.uni-bonn.de (PK); sambaran@astro.uni-bonn.de (SB)}
Pavel Kroupa$^2$\footnotemark[1] and Sambaran Banerjee$^1$\footnotemark[1]
\\
$^1$Argelander-Institut f\"ur Astronomie, Auf dem H\"ugel 71, D-53121 Bonn, Germany\\
$^2$Helmholtz-Institut f\"ur Strahlen- und Kernphysik (HISKP), University of Bonn, Nussallee 14-16, 
D-53115 Bonn, Germany}

\begin{document}

\date{Accepted 2013 November 12.  Received 2013 November 11; in original form 2013 October 10}
\maketitle

\begin{abstract}
R144 is a recently confirmed very massive, spectroscopic binary which  
appears isolated from the core of the massive young star cluster R136.  
The dynamical ejection hypothesis as an origin for its location is claimed improbable by Sana et al. 
due to its binary nature and high mass. 
We demonstrate here by means of direct N-body calculations that a very massive binary system can 
be readily dynamically ejected from a R136-like cluster, through a close encounter with a very massive system. 
One out of four N-body cluster models produces a dynamically ejected very massive binary system 
with a mass comparable to R144. 
The system has a system mass of $\approx 355~\msun$ and is located at 36.8~pc from the centre 
of its parent cluster, moving away from the cluster with a velocity of $57~\kms$ at 2~Myr 
as a result of a binary-binary interaction.
This implies that R144 could have been ejected from R136 through a strong encounter with an other 
massive binary or single star. In addition, we discuss all massive binaries and single stars which 
are ejected dynamically from their parent cluster in the N-body models.  
\end{abstract}


\section{introduction}
The formation process of massive stars is still poorly understood. Among many issues, 
whether massive stars form only in clusters or whether they can form in isolation has been a major debate to this day. 
\citet{Wet05} suggested that about 4 per cent of the Galactic field O stars are formed in isolation. 
Later, every one of these candidates for isolated massive star formation were found to be runaways 
\citep[and references therein]{Get12} or to be consistent with having been generated via the two-step ejection
process described by \citet{PK10}. 
This indicates that probably all O stars in the Galactic field likely originate from star clusters \citep{Get12}.

Isolated massive stars have been found not only in the Galaxy but also in our nearest neighbour galaxies, 
the Magellanic Clouds \citep[and references therein]{Bret12,Get12,Oet13}.
Even stars more massive than $100~\msun$ are found in apparent isolation in the 30~Dor region 
in the Large Magellanic Cloud (LMC), among them VFTS~682 \citep{Bet11} and R144 \citep{Set13}. 
The presence of these massive stars in isolation has been interpreted by the respective authors 
to mean that very massive stars can form in isolation.
Although dynamical ejection as the origin of VFTS~682 was claimed challenging due to its high mass 
($\approx 150~\msun$) \citep{Bet11}, with the direct N-body calculations of R136-like clusters \citet*{BKO12} 
showed that the star would have been dynamically ejected from R136. 
The dynamical ejection process can thus explain these observed very massive isolated stars, suggesting there is no need 
to postulate isolated massive star formation.
In this study, our main interest is R144, a recently confirmed very massive spectroscopic binary which is located 
at a projected distance of 59~pc from R136 \citep{Set13}. \citet{Set13} argued that the ejection from R136 
via a dynamical interaction is improbable. It is thus important to investigate whether a very massive binary, 
such as R144, can be dynamically ejected from a cluster like R136, in order to prove that the system 
can originate only from in situ formation.

Using three-body scattering experiments, \citet{GG11} showed that a very massive star 
can attain a high velocity through a close encounter with a very massive binary such as WR20a and NGC3603-A1.  
They also suggested that the ejection of a very massive star can accompany a recoiled very massive binary. 
However, so far there has been no study to find a very massive binary ejected dynamically from its parent cluster 
with full N-body calculations for a realistic cluster model comparable to R136. 
It is only recently that massive young star cluster models comparable to R136 have been computed 
with the direct N-body method \citep{FP11,BKO12,FSP12}. Among them, \citet{BKO12} constitutes the only 
theoretical dataset which initially applied a realistic high fraction of massive close binaries, including
ones with primary masses larger than $100~\msun$. Observed O-star binaries in the Tarantula Nebula, in which R136 is at the centre, 
are found to favour short binary orbital periods \citep{Set13a}.  
\citet{BKO12} performed direct N-body calculations for R136-like cluster models with 
primordial massive binaries but mainly focused on ejected massive single stars which 
have similar properties as VFTS~682.  
Here we revisit the N-body data to study if a very massive binary such as R144 in the LMC could be ejected from its birth cluster 
via a close encounter with other cluster members. We would like to stress that the models used here are the exact same 
ones as used in \citet{BKO12} and that they were not generated with the specific purpose of addressing the problem at hand 
(can R144-type objects be ejected?).

We briefly describe the N-body models of \citet{BKO12} in Section~2. A dynamically ejected very massive binary
found in the N-body calculations is described and compared to R144 in Section~3.
Section~4 presents dynamically ejected massive binaries and single stars from the N-body calculations.
The conclusion is presented in Section~5.

\section{N-body Models}
We revisit the direct N-body calculations for a R136-like cluster performed by \citet{BKO12} to study if 
a very massive binary, as massive as R144, can be dynamically ejected. The models used here are thus not
specially made to study the ejection of binaries. The models were integrated with 
the Aarseth {\sevensize NBODY}6 code \citep{Aa99,Aa03}.

The initial cluster mass is set as $10^{5}~\msun$ with 170667 stars. 
The model clusters have initial half-mass radii, $r_{\rmn{h}}$, of 0.8~pc. 
The initial positions and velocities of stellar systems, single stars or centre-of-mass of binary systems, are generated 
following the Plummer density profile and assuming the cluster to be in virial equilibrium.   
Individual stellar masses are randomly derived from the two-part power law canonical initial mass function (IMF) 
with power indices of $\alpha_{1}=1.3$ for $0.08~\msun \le m < 0.5~\msun$ 
and of $\alpha_{2}=2.3$ for $m \ge 0.5~\msun$ \citep{Ket13} in a stellar mass range $0.08~\msun \le m \le 150~\msun$.
The model clusters are initially fully mass-segregated by assigning more massive stars 
to be more bound to the cluster \citep*{BDK08}.  Due to the high computational cost, 
only stars more massive than $5~\msun$, i.e. $\approx$ 2000 stars, are initially 
in a binary system ($\approx1000$ binary systems).
The exact number of binaries slightly differs from model to model as it is determined by the number 
of stars more massive than 5~$\msun$ which fluctuates among the models.  
For a realistic initial setup of massive binaries, for them to be consistent with their observed 
orbital parameters \citep{Set12,Set13a}, a star is paired with the next less massive star 
and periods are generated using a uniform distribution over a period
range between $10$ and $10^{4}$~days for binaries with a primary more massive than $20~\msun$, 
and using the \citet{PK95} distribution for lower mass binaries. 
The thermal distribution was used for the initial eccentricity distribution of binary orbits. 
Stellar evolution was included and stellar collisions were allowed.  
Four calculations with initial configurations generated by different random seed numbers were carried out. 
More details can be found in \citet{BKO12}.  

In the following sections we discuss results from the snapshots of the N-body integrations at 2~Myr. 
It is worthy to mention that by 2~Myr the total number of stars is reduced by 30-35 due to stellar collisions
and that most of the collisions took place at the very beginning of the calculations due to 
initially short periods and high eccentricities of massive binaries \citep*{BKO12b}.

\section{A very massive binary dynamically ejected via a binary-binary interaction}
Our motivation for this study is the recently confirmed massive binary star R144 which is found 
about 60~pc away from R136 in projection and has an estimated mass of $200-300~\msun$ \citep{Set13}.  
Apparently not being inside of any star cluster and its high mass are interpreted such 
that the system is very unlikely to have been ejected via a close encounter and thus that it is a good candidate 
of isolated massive star formation \citep{Set13}. 
To check this argument, we study if such a system can appear in the direct N-body calculations of 
realistic R136-like cluster models.   

We find that one very massive binary system with a system mass, $m_{\rmn{sys}} = 355.9~\msun$,
has been dynamically ejected and is located 36.8~pc away from the centre of its birth cluster moving 
with a velocity of $57~\kms$ at 2~Myr (system ID 4-5 in Table~1).\footnote{Note that the binary system was 
misclassified as two single stars in \citet{BKO12}.} 

Theoretically, after an energetic encounter between a binary and a single star, the recoil of the very massive binary 
can be accompanied by an ejection of a very massive star because of momentum conservation \citep{GG11}.
However, it was not certain whether such an event can occur in a cluster. 
Here we show that a binary system even more massive than R144 can be ejected via a few-body close encounter 
in a R136-type cluster.
 
The above binary is composed of two stars with present masses of $244.0~\msun$ and $111.9~\msun$. 
The primary of the binary with a present mass of $244.0~\msun$ is, in fact, a merger product which was initially 
a binary with initial component masses of $134.9~\msun$ and $129.2~\msun$. The secondary star was initially in 
a binary with an ejected single star 4-10 in Table~2 which is found on the opposite side of the cluster 
with respect to the binary system.  

The orbital period of the binary found in the N-body model is 1955.7~days with a highly eccentric orbit 
(eccentricity, $e = 0.78$). 
The period estimates of R144 from observations are between two and six months or can be up to one year 
if the orbit is highly eccentric \citep{Set13}. 
Even though the very massive, ejected binary found in the N-body model has a relatively long period, 
a massive binary with a period as short as 7.7 days can also readily be dynamically ejected (Section~4 and Table~1).  

\begin{figure*}
\includegraphics[width=80mm]{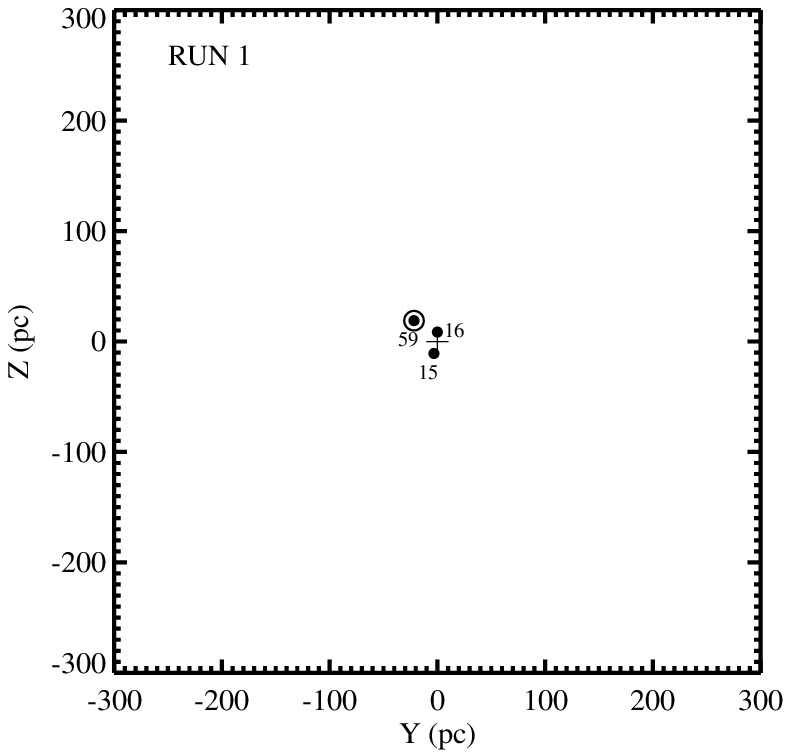}
\includegraphics[width=80mm]{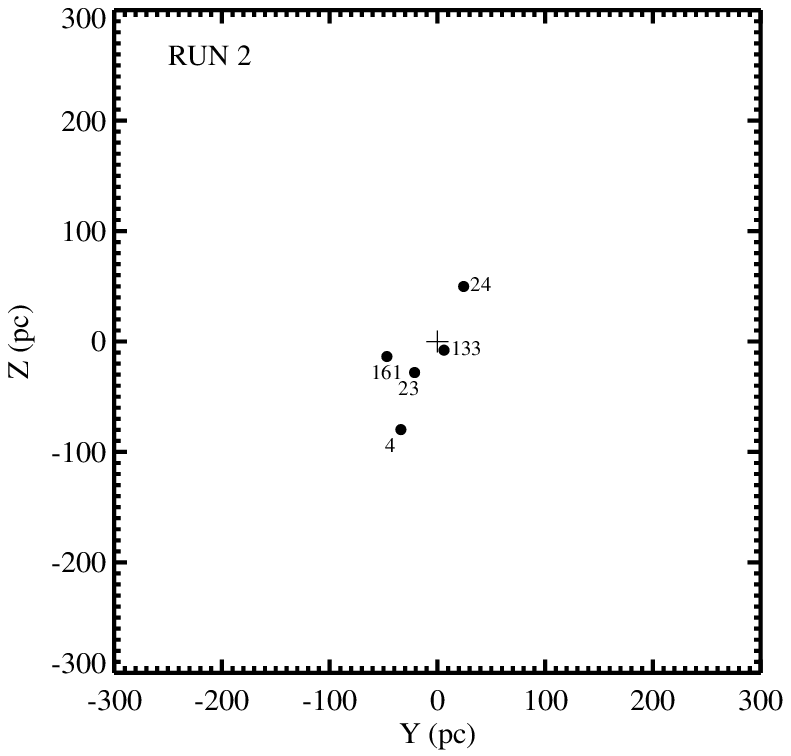}\\
\includegraphics[width=80mm]{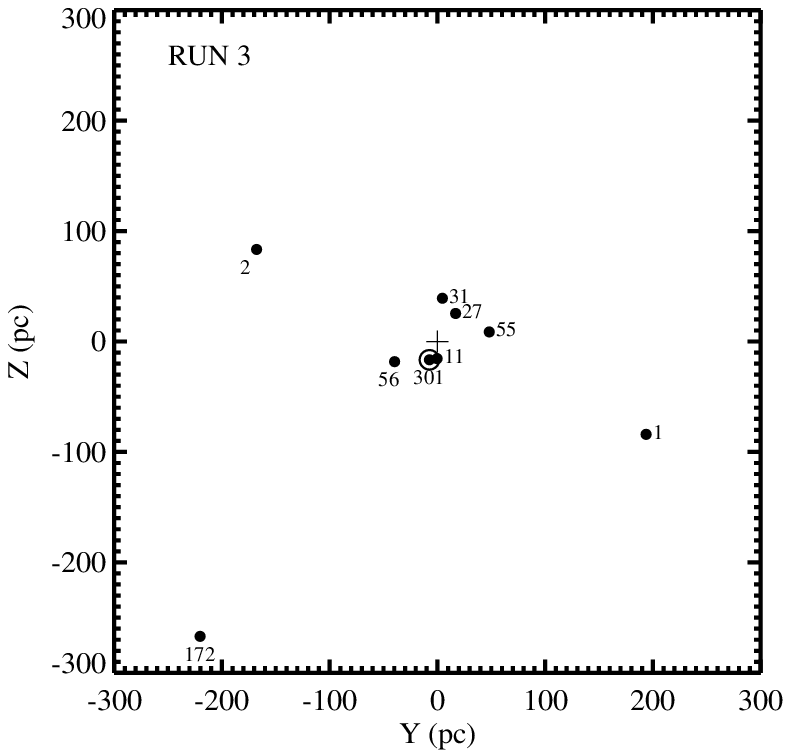}
\includegraphics[width=80mm]{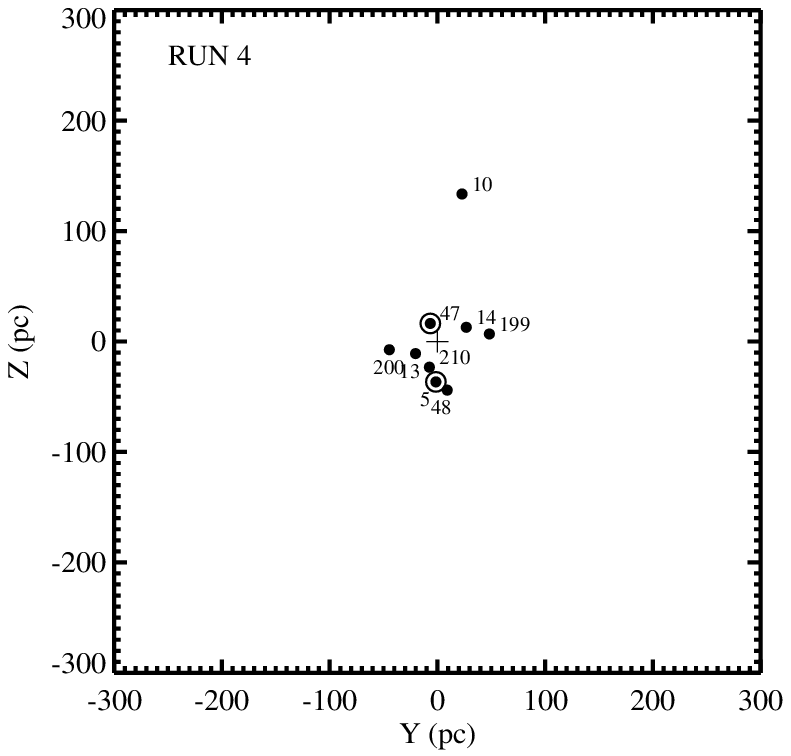}
\caption{Snapshots of the ejected massive stars for each run projected on the YZ plane at 2~Myr.
 Dots are single stars and dots with an outer circle are binary systems. Cluster centre is indicated
 with a cross in each panel. System identification numbers are noted.}
\label{sshots}
\end{figure*}

\citet{GG11} suggested that R145, another very massive binary ($m_{\rmn{sys}} \approx 240~\msun$, 
Chen\'e et al. 2010, period $\approx 159$~days, Schnurr et al. 2009) 
located outside of the R136 core (at a projected distance $\approx 19$~pc), could be such a recoiled 
very massive binary having been involved in an energetic encounter which would have ejected 
the runaway B2 star Sk--69\degr206. In the case of the N-body models here, there is a single massive star ejected
in the opposite direction of the ejected binary (star 4-10 in Table~2 and see the panel RUN4 in Fig.~\ref{sshots}).
If R144 is such a case as well, then one may find an ejected massive star on the opposite side of its birth cluster.

\begin{table*}
\caption{
 List of massive binaries ejected from all four model-cluster integrations by 2~Myr.
 ID in column~1 indicates identification of run and the system. System mass, $m_{\rmn{sys}}$,
 distance from the cluster center $r$, and 3D velocity, $v$, of the systems are listed in columns 2-4.
 Furthermore, columns 5-10 list parameters of binaries such as  primary- and secondary mass
 ($m_{1}$ and $m_{2}$, respectively), orbital period ($P$), semi-major axis ($a$),
 eccentricity ($e$) and orbital velocity ($v_\rmn{orb}$)
 when a circular orbit is assumed. Ratio of binding energy of the binaries (equation~\ref{ebin})
 to the initial total energy of the cluster (equation~\ref{etot}) is listed in column~11.
 All values are from snapshots at 2~Myr.
}
\begin{tabular}{@{}lcccccccccc@{}}
\hline
ID&$m_{\rmn{sys}}$&$r$&$v$&$m_{1}$&$m_{2}$&$P$&$a$&$e$& $v_\rmn{orb}$& $E_{\rmn{b}}/E_{\rmn{tot,ecl}}$ \\
  &($\msun$)  & (pc) & ($\kms$) & ($\msun$)  & ($\msun$) & (days) & (au) & &($\kms$) & \\
\hline
1-59$^{a}$  &110.8  &29.1 &43.8 &55.5   &55.3   &19.4    &0.68   &0.44 &381.0  & 0.19 \\
3-301 &40.6   &20.9 &11.1 &20.3   &20.3   &7.7     &0.26   &0.57 &371.0  & 0.07 \\
4-5   &355.9  &36.8 &57.0 &244.0  &111.9  &1955.7  &21.69  &0.78 &120.8  & 0.05 \\
4-47  &108.9  &29.1 &27.5 &60.2   &48.6   &35.1    &1.00   &0.86 &310.9  & 0.13 \\
\hline
\end{tabular}
\medskip\\
$^{a}$ The initial orbital parameters of the system are $P=30.7$~d, $a=0.93$~au, and $e=0.53$.
\end{table*}

\section{Dynamically ejected massive systems}
In this study, we define an ejected massive system as a system whose current (primary, in case of a binary) mass is 
larger than $20~\msun$ and whose distance from the parent cluster density centre is greater than 10~pc at 2~Myr.
A 'system' refers to either a single star or a binary. 
We find 26 ejected massive systems from all four calculations. 
4 systems are binaries (Table~1) and the other 22 systems are single stars (Table~2).    

Among all four dynamically ejected binary systems, three have system masses larger than $100~\msun$ 
including the one discussed in the previous section. Furthermore, about half (10) of the single stars are more massive 
than $90~\msun$. This indicates that very massive stars can be efficiently ejected from a massive young star cluster 
within the first few Myr of cluster evolution (see figure~4 in Banerjee et al. 2012) since those very massive stars 
are formed in or migrate into \citep*{Bet04} the cluster core dynamically at a very early age of the cluster, 
given the high densities of star-burst clusters such as R136.

\begin{table}
\caption{List of the ejected massive single stars by 2~Myr. Same as columns 1-4 in Table~1.}
\begin{tabular}{@{}lccc@{}}
\hline
ID&$m_{\rmn{sys}}$&$r$&$v$\\
 &  ($\msun$)  & (pc) & ($\kms$)\\
\hline
1-15   & 90.9   &12.6   &73.3\\
1-16   & 90.6   &10.7   &59.9\\
2-4    & 129.9  &87.9   &69.7\\
2-23   & 69.3   &35.8   &38.8\\
2-24   & 67.9   &62.0   &69.0\\
2-133  & 33.6   &10.4   &9.3 \\
2-161  & 49.0   &53.6   &72.8\\
3-1    & 132.4  &211.4  &131.8\\
3-2    & 129.0  &187.4  &117.0\\
3-11   & 116.9  &36.9   &32.8\\
3-27   & 169.1  &30.8   &28.9\\
3-31   & 74.3   &40.4   &75.0\\
3-55   & 61.3   &49.5   &39.9\\
3-56   & 61.2   &44.1   &35.0\\
3-172  & 29.2   &348.3  &343.0\\
4-10   & 111.2  &137.3  &215.8\\
4-13   & 106.5  &32.1   &50.5\\
4-14   & 105.9  &30.8   &48.1\\
4-48   & 59.1   &57.1   &55.6\\
4-199  & 26.1   &48.9   &55.2\\
4-200  & 26.1   &46.8   &53.2\\
4-210  & 25.3   &24.5   &93.9\\
\hline
\end{tabular}
\end{table}

\begin{figure}
\includegraphics[width=85mm]{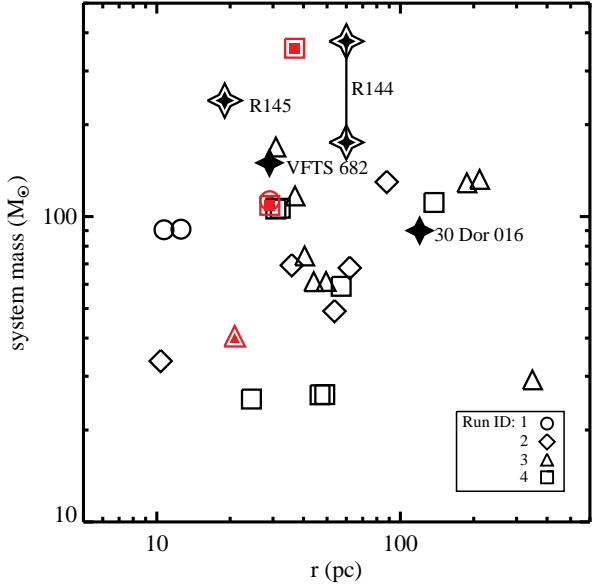}
\caption{Distance from the cluster centre vs. system mass of the ejected massive systems.
 Single stars are black open symbols. Red symbols with small filled symbols inside are binary systems. 
 Star symbols are the {\it observed} massive systems located
 outside of the R136 core, e.g. R144 \citep{Set13}, R145 \citep{Set09,Cet10}, 30 Dor 016 \citep{Eet10}, 
 and VFTS 682 \citep{Bet11}. The filled stars are single stars while open stars with small filled 
 symbols confirmed binary systems.
 The two mass estimates of R144 are connected with a solid line. As is evident, the theoretical systems 
 cover the range of observed cases.}
\label{mrfig}
\end{figure}

\begin{figure}
\includegraphics[width=85mm]{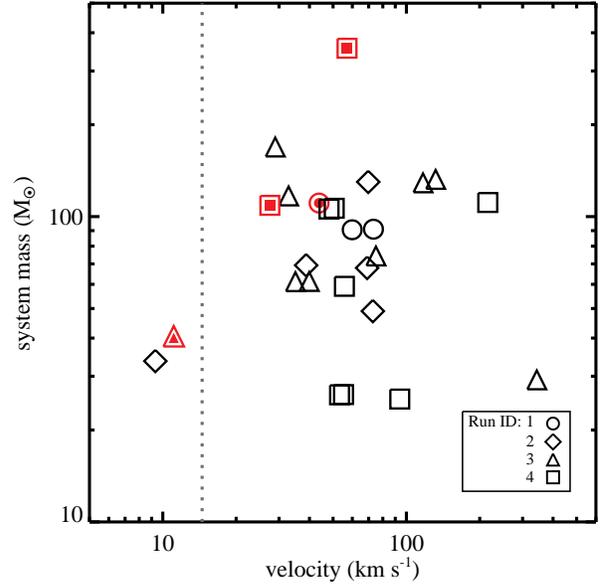}
\caption{Velocity of the ejected massive system with respect to the cluster centre vs. system mass of the ejected massive systems.
 Symbols are the same as in Fig.~\ref{mrfig}.  The dotted line represents the initial rms velocity 
dispersion of the model clusters,  $\approx 14.5~\kms$.}
\label{mvfig}
\end{figure}

In the models studied here 
the binaries are ejected with relatively lower velocities and are found relatively close ($r < 38$~pc) 
to the clusters compared to the ejected single stars which are found up to about 350~pc away from 
their birth clusters at 2~Myr (Figs \ref{mrfig} and \ref{mvfig}). 
A few single stars more massive than $100~\msun$ are ejected with a velocity greater than $100~\kms$ up to $\approx200~\kms$. 
Especially, one star with a mass of $111.2~\msun$ moves away from its parent cluster with a velocity of $215.8~\kms$. 
This was probably ejected by the same binary-binary interaction which has ejected the very massive binary described in Section~3; 
it is located on the opposite side of the cluster.  

As shown in Fig.~\ref{mrfig}, the ejected massive systems in the models cover the range of observed very 
massive systems outside of the R136 core in the system-mass -- distance from the cluster centre space. 
The Jacobi radius of the cluster with a mass of $10^{5}~\msun$ at the location of R136 in the LMC, 
$\approx 1$~kpc in projection relative to the \mbox{H\,{\sc i}} rotation centre of the LMC \citep{MG03}, 
would be $\approx 43$~ pc \citep{Bet13},\footnote{Tidal field is not included in the N-body calculations.} 
so some of the ejected massive systems are still located inside the Jacobi radius and may be under 
the influence of the cluster potential.
However, all ejected massive systems are unbound to the cluster, 
i.e. their velocities exceed the escape velocity at their location (e.g. initial escape velocity at 10 pc 
of the model clusters $\approx 9.5 \kms$).   
Therefore most of those still remaining inside the Jacobi radius at 2~Myr will move outside the Jacobi radius 
within less than 0.5~Myr considering their high velocity. 

Like the distances from the cluster centre, the velocities of the ejected massive systems 
are also spread over a large range, from 9 to 350~$\kms$.  
There seems no evident correlation between the system mass and the velocity of ejected massive systems 
(Fig.~\ref{mvfig}) though there is a little tendency that more massive systems have on average
higher velocities \citep{BKO12}. The initial rms velocity dispersion of the model clusters, 
$\approx 14.5~\kms$ (equation 8.69 in Kroupa 2008), is indicated in Fig.~\ref{mvfig}. Most of the ejected systems 
have attained velocities significantly larger than the rms velocity dispersion. 

\begin{figure}
\includegraphics[width=80mm]{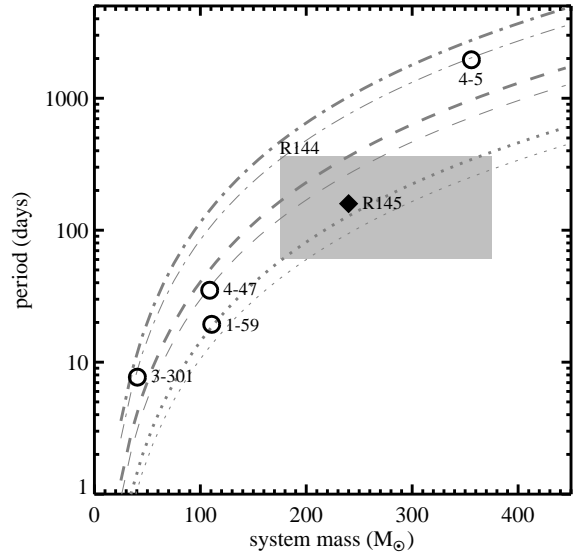}
\caption{System mass vs. period of the ejected binary systems. Open circles are the binaries from the N-body models 
 while the filled diamond is R145 \citep{Set09,Cet10}. A large grey box indicates the range of the values of R144 in \citet{Set13}. 
 Curves present equal binding energies, i.e. 5 (dot-dashed), 10 (dashed), 20 (dotted) per cent of the total energy of the cluster 
 (equation~\ref{etot}) assuming a certain mass-ratio (thick lines, $q=1$; thin lines, $q=0.4$).}
\label{mpfig}
\end{figure}

One of the interesting findings here is that all ejected binaries but one have a short period 
($P\lesssim35$~days, Fig.~\ref{mpfig}). This may mainly result from the initial period distribution which 
the N-body models adopted. Indeed the two shortest period binaries are natal pairs (1-59 and 3-301 in Table~1). 
However, even a binary which has dynamically formed probably after a binary-binary interaction (4-47\footnote{This binary 
system is also not found in a list of ejected stars in \citet{BKO12} because the system merges 
before 3~Myr due to its highly eccentric orbit.} in Table~1) 
also has a short period, $P=35.1$~days. 
We expect that ejected massive binaries from R136 would preferentially have short periods like the result presented here, 
because the observed period distribution of O-type star binaries in the Tarantula Nebula, which hosts the cluster R136 
at its centre, favours short periods \citep{Set13a}, in fact even shorter ones than used in the N-body models here.  
Furthermore, an energetic encounter between a binary and another system can result in the shrinkage of the binary's orbit.
Indeed the orbit of one ejected binary, 1-59, has shrunk compared to its initial one: the period has changed from 30.7 to 19.4~d, the semi-major axis from 0.93 to 0.68~au, and the eccentricity from 0.51 to 0.44.  
Note that the other natal binary, 3-301, keeps its initial binary orbital parameters 
implying that the binary was treated as a single star in the interaction which has caused its ejection. 

By fixing the binding energy, $E_{\rmn{b}}$, and mass-ratio of a binary, 
$q=m_{2}/m_{1}$ where $m_{1}$ and $m_{2}$ are masses of the binary components in $\msun$ and $m_{1}>m_{2}$, 
the binary period can be expressed as a function of system masses,
 \begin{equation}
 P_{\rmn{yr}} = \left[ -\frac {206264.8 G q}{2(1+q)^2 E_{\rmn{b}}} \right]^{\frac{3}{2}} m_{\rmn{sys}}^{\frac{5}{2}}, 
 \end{equation}
 where $P_{\rmn{yr}}$ is the period in years, $G$ is the gravitational constant ($G=0.0045~\rmn{pc}^{3}~\msun^{-1}~\rmn{Myr}^{-2}$)  
 and the binding energy of the binary in units of $\msun~\rmn{pc}^{2}~\rmn{Myr}^{-2}$ is, 
 \begin{equation}\label{ebin}
 E_{\rmn{b}} = -\frac{Gm_{1}m_{2}}{2 a_{\rmn{pc}}}=-\frac{G q}{2(1+q)^2}\frac{m_{\rmn{sys}}^2}{a_{\rmn{pc}}},
 \end{equation}
where $a_{\rmn{pc}}= a_{\rmn{au}}/206264.8$~pc and $a_{\rmn{au}}$ is the semi-major axis in au.
Curves in Fig.~\ref{mpfig} indicate 5, 10 and 20 per cent of initial total binding energy of the cluster \citep{HH03}, 
 \begin{equation}\label{etot}
 E_{\rmn{tot,ecl}}= -\frac{3 \pi }{32} \frac{G M_{\rmn{ecl}}^{2}}{r_{\rmn{P}}},
 \end{equation}
in units of $\msun~\rmn{pc}^{2}~\rmn{Myr}^{-2}$ where $r_{\rmn{P}}$ is the Plummer radius $\approx0.766 r_{\rmn{h}}$ in units of pc. 
The figure indicates a possible correlation between system mass and a period of the ejected massive binaries, but the number of 
cases is too small to be conclusive.
The ejected binaries have binding energies between 5 and 20 per cent of the initial total energy of the cluster. 
They thus contain a significant amount of energy. 

As stellar identification numbers in the {\sevensize NBODY}6 code are consecutive for initial binary components, 
we can identify if the ejected stars have been initially paired with other ejected stars. 
Among the ejected single stars, six natal pairs 
(i.e. 12 stars) can be deduced. Furthermore, two of the ejected single stars, 4-10 and 4-48, were a companion 
of the two ejected binaries, 4-5 and 4-47 respectively, which have exchanged their companions.   
Both components of one binary are shot out into opposite directions (Fig.~\ref{sshots}) almost conserving 
momentum. However there should be a third body, likely an other binary system, to operate the ejection process, 
thus the momentum of one ejected star is not exactly identical to the other's. 
In addition, dynamically ejected systems are still under the influence of the gravitational potential 
of the cluster right after ejection while they move away from the cluster. 
This can reduce the velocity of the systems, especially of massive ones with a low velocity. 
Associating isolated massive stars with particular ejection events is thus not straightforward.

\section{Conclusion}
We find a very massive binary ($m_{\rmn{sys}} \approx 355.9~\msun$) with a period $P\approx 2000$~d 
which is dynamically ejected via a binary-binary interaction in previously published direct N-body models of a R136-like cluster. 
This suggests that a very massive binary system such as R144 can be dynamically ejected via a close encounter
with an other very massive binary or single star and that they then appear in isolation.
Three other massive binaries are found being ejected in the four R136-model clusters studied here. 
Among them two have a system mass larger than $100~\msun$.
Three out of four binaries have a period shorter than $\approx$ 35 days. 
From this result and the observed period distribution of O-type star binaries in the Tarantula Nebula \citep{Set13a}, 
we expect that ejected massive binaries from R136 would preferentially have short periods contrary 
to the discussion in \citet{Set13}.   

We show that very massive stars can be ejected with a high velocity via a three- or four-body interaction 
in a realistic full-cluster size N-body calculation, even a star more massive than $100~\msun$ can obtain a velocity 
of more than $200~\kms$. The ejected massive stars travel up to 350~pc away from their birth clusters 
within the first 2~Myr of their cluster evolution.

Among the 26 dynamically ejected massive systems, 15 systems ($\approx 58$~per cent) are more massive 
than $90~\msun$. Thus in the first few Myr of evolution of a massive cluster like R136, mostly very massive stars are dynamically 
ejected since they are the ones located at the very centre of the cluster where the probability for close encounters is the highest. 
Massive young star clusters such as R136 would mainly contribute very massive stars into the field 
during the first few Myr of their evolution. 
Correcting the measured stellar mass function within R136 for the likely massive stars lost from the cluster through 
ejections reveals a flattening of the IMF at high stellar masses \citep{BK12}.

Thus, more than one very massive binary can be ejected from a massive young star cluster by dynamical interactions. 
One out of four runs produced two very massive binaries ejected dynamically by 2~Myr.
Thus the cluster R136 probably produced both R144 and R145 through close encounters with other very massive systems. 
In particular, the isolated formation scenario of R144 is unnecessary.

\end{document}